# Aggregated journal-journal citation relations in Scopus and Web-of-Science matched and compared in terms of networks, maps, and interactive overlays

Loet Leydesdorff,[a] Félix de Moya-Anegón,[b] & Wouter de Nooy [c]

**Abstract**

We compare the network of aggregated journal-journal citation relations provided by the Journal Citation Reports (JCR) 2012 of the Science and Social Science Citation Indexes (SCI and SSCI) with similar data based on Scopus 2012. First, global maps were developed for the two sets separately; sets of documents can then be compared using overlays to both maps. Using fuzzy-string matching and ISSN numbers, we were able to match 10,524 journal names between the two sets; that is, 96.4% of the 10,936 journals contained in JCR or 51.2% of the 20,554 journals covered by Scopus. Network analysis was then pursued on the set of journals shared between the two databases and the two sets of unique journals. Citations among the shared journals are more comprehensively covered in JCR than Scopus, so the network in JCR is denser and more connected than in Scopus. The ranking of shared journals in terms of indegree (that is, numbers of citing journals) or total citations is similar in both databases overall (Spearman's $\rho > 0.97$), but some individual journals rank very differently. Journals that are unique to Scopus seem to be less important—they are citing shared journals rather than being cited by them—but the humanities are covered better in Scopus than in JCR.

**Keywords:** journal, map, matrix, network, Scopus, WoS, JCR

[a] University of Amsterdam, Amsterdam School of Communication Research (ASCoR), Kloveniersburgwal 48, 1012 CX Amsterdam, The Netherlands; loet@leydesdorff.net ; * corresponding author
[b] CSIC, SCImago Research Group, Centro de Ciencias Sociales y Humanas, Instituto de Políticas y Bienes Públicos, Calle Albasanz 26, Madrid 28037, Spain; felix.moya@scimago.es .
[c] University of Amsterdam, Amsterdam School of Communication Research (ASCoR); w.denooy@uva.nl .

## 1. Introduction

In this study, we compare the networks of aggregated journal-journal citation relations as provided by the Journal Citation Reports (JCR) 2012 of the Science and Social Science Citation Indexes (SCI and SSCI) with similar data for 2012 based on Scopus. JCRs have been compiled from the citation indices since 1975 for SCI and since 1978 for SSCI, first in print and microfiches, but since 1994 also in electronic format (Garfield, 1972; 1979). JCR data are brought online by Thomson Reuters at the Web of Science (WoS) for the period 1997-2012. This interface is well known for providing also impact factors and similar indicators of journal status (e.g., Bergstrom, 2007; Price, 1970).

Elsevier's Scopus has provided an alternative to WoS since 2004. Scopus contains reliable data since 1996 (Ove Kahler, *personal communication*, August 28, 2009). This data is used, among other things, for the Scimago journal rankings (SJR) at http://www.scimagojr.com/journalrank.php (Gonzalez-Pereira, de Moya-Anegón & Guerrero, 2010; Guerrero-Bote & Moya-Anegón, 2012) and the journal indicator SNIP[1] provided by Elsevier in collaboration with the Center for Science and Technology Studies (CWTS) in Leiden at http://www.journalmetrics.com (Moed, 2010; Waltman *et al.*, 2013). These two journal indicators in Scopus assume 1999 as the first year of the analysis. A third database that provides citation data is Google Scholar (GS), but GS data cannot be aggregated at the journal level across the file because of technical limitations to the accessibility of this data online (Jacsó, 2012).[2]

[1] SNIP is an abbreviation for "Source-Normalized Impact per Paper." However, SNIP is a journal indicator in Scopus comparable to the Impact Factor in JCR (Moed, 2010).
[2] Google Scholar has a limit of 1,000 records that can be downloaded from the results of a search. In Scopus, this number is 2,000, but Scopus is also commercially available as a database for further analysis.

These three databases have been compared in terms of different characteristics (e.g., Bar-Ilan, 2008; Meho & Rogers, 2008), but mostly in terms of their use for evaluative bibliometrics (e.g., Franceschet, 2010) both at the levels of institutions and journals (e.g., López-Illescas *et al.*, 2008). Using both WoS and Scopus as entire databases off-line, Archambault *et al.* (2009) found correlations larger than .99 for country shares in terms of numbers of publications and citations. Because Scopus covers approximately twice as many journals as WoS, one would expect absolute scores to be higher in Scopus than in WoS.

In addition to hierarchical rankings, the network of citations among journals contains structural information that can be used for mapping purposes. Mapping JCR data has a long tradition in scientometrics. Groupings of journals in the citation network suggest a way to demarcate specialties (Rafols & Leydesdorff, 2009). In the mid-1980s, several research teams almost simultaneously began to use visualization techniques such as multi-dimensional scaling (MDS) for generating journal maps from the aggregated citations among journals (Doreian, 1986; Doreian & Fararo, 1985; Leydesdorff, 1986; Tijssen, de Leeuw & Van Raan, 1987). However, these maps were local journal maps of limited sets, that is, of the order of between ten and one hundred journals.

Beyond local maps, the advent of visualization techniques in Windows during the 1990s (e.g., the graphical user interface) and computers with more memory has made it feasible since the early 2000s to develop global maps—that is, including all the available data about journal-journal citations in a single map (e.g., Boyack, Klavans & Börner, 2005; de Moya-Anegón *et al.*, 2004; Rosvall & Bergstrom, 2008). On the basis of a number of studies Klavans & Boyack (2009) concluded that in the meantime a consensus about the structure of the journal literature had formed from the various mapping efforts: the maps indicate a torus-like shape in which the disciplines are grouped in a circle with the computer sciences relatively more toward the center (*ibid.*, p. 471). However, the classification of journals using one computer routine or another remains uncertain if one zooms in to a finer-grained level (Leydesdorff, 2006; Rafols & Leydesdorff, 2009; Zitt *et al.*, 2005). Bollen & Van Sompel (2009) added that by using clickstream data, the humanities could fill important holes in this record.

Unlike JCR data—that do not cover the *Arts & Humanities Citation Index* (A&HCI; cf. Leydesdorff, Hammarfeld, and Salah, 2011)—Scopus includes data for the humanities. Leydesdorff, de Moya-Anegón & Guerrero-Bote (2010) generated local maps on the basis of Scopus data of 2008. Recently, a global map with the facility to overlay data from Scopus for the period 1996-2012, was made available for interactive usage at http://www.leydesdorff.net/scopus_ovl (Leydesdorff, de Moya-Anegón & Guerrero-Bote, in press). In the latter study, we developed a system equivalent to the interactive overlay maps that were generated for JCR data from 2011 (Chen & Leydesdorff, 2014; Leydesdorff, Rafols, & Chen, 2013).

In the case of Scopus data, we decided at the time to use the aggregated set of 1996-2012 because (*i*) Scopus data is differently organized—that is, not on a yearly basis like JCR—and (*ii*) the set of journals included in Scopus contained 20,554 journals (Oct. 2013), whereas combined JCR 2012 data covered only 10,936 journals. The more peripheral journals added to Scopus can be expected to make the network so sparsely populated that one risks losing a perspective on structure because of isolates and potentially isolated islands. In other words, the aggregation over more than a decade is expected to make the basemap more reliable.

Acknowledging a suggestion by one of the referees of this previous study, we follow up in this study using exclusively 2012 citation relations in the Scopus database for a systematic comparison with JCR 2012 data of WoS, both in terms of global maps and network characteristics. The definition of "2012 data," however, may differ somewhat for each of the two database providers. JCR uses a cut-off date in March, when the previous year is assumed to be finished (Marie McVeigh, *personal communication*, 7 April 2010), in order to produce a beta-version of JCR for the previous year during the summer (June/July). In October, this is followed up with a second (error-corrected) version. The production of the Scopus database is a continuous operation; the 2012 data was extracted by one of us on October 13, 2013. The SCImago group that processes Scopus data since 2009 has the experience that changes after October do not affect the database of the previous year any longer seriously.

The citation matrix of each of the databases is an asymmetrical matrix ("cited" *versus* "citing") with journal self-citations on the main diagonal. For the visualizations, we limit ourselves to the citing side. "Citing" is a variable that evolves as action over the years, whereas "cited" represents an accumulation of previous years into a shared knowledge base. The network analysis, however, will be pursued from the "cited" side because when comparing journals bibliometrically, one is interested in their citation impact during the year(s) under study.

Furthermore, network analysis uses a semantics different from bibliometrics: "citedness," for example, can be measured as "indegree", that is, in terms of the number of journals that are citing the journal under study. "Weighted indegree" would then add up to "total cites," but the initial assumption in (unweighted) network analysis is that each link is either ON or OFF. Measures such as "indegree centralization" for the network are not properly defined for weighted networks. Degree centralization of a network, for example, is defined as the variation in the degrees (that is, number of links) of vertices (nodes) divided by the maximum degree variation that is possible in a network of the same size (De Nooy *et al.*, 2011: 144).

In summary, scientometric visualization and network analysis provide different perspectives on the data. Whereas we use the citation matrix for the network analysis, the mapping requires normalization because otherwise the major journals and their relations could be expected to overshadow the smaller (more specialist) journals on the map (see, e.g., Figure 2 in Leydesdorff *et al.*, in press). The cosine is used for normalization because citation distributions are non-normal (Ahlgren *et al.*, 2003); VOSviewer (Van Eck & Waltman, 2010) will be used for the clustering and visualization of this data.[3]

## 2. Data

The data for Scopus 2012 was extracted from the Scopus database (1996-2012) by one of us in October 2013 in the context of another project (Leydesdorff, de Moya-Anegón & Guerrero-Bote, in press). The number of source journals in the database (Scopus) was 20,554 at the time, of which 20,172 (98.1%) were involved in citation relations during 2012. Of the more than 400 million possible links among these 20,000+ units of analysis, 6,672,033 (1.6%) are realized, or—in the terminology of social network analysis—the density is 0.0164. However, 3,983,302 of these links are single or invalid citation relations.[4] Since single citations are aggregated in the JCR under “All others,” [5] we discarded these values and pursued the analysis with the 2,688,731 remaining links which contain 36,748,156 citation relations.

[3] VOSviewer is a freeware program for network visualization available at http://www.vosviewer.com .

**Table 1:** Descriptive statistics of the data.

| | **Scopus 2012** | **JCR 2012 (SCI + SSCI)** | **JCR SCI** | **JCR SSCI** |
|---|---|---|---|---|
| ***N* of journals** | 20,172 * | 10,936 | 8,471 | 3,047 |
| **Citation links** | (6,672,033) | | | |
| | 2,688,731 ** | 2,350,491 | 2,122,083 | 253,320 |
| **Sum of citations** | (40,731,458) | | | |
| | 36,748,156 ** | 37,759,948 | 35,721,660 | 2,454,015 |
| **Self-citations** | 2,898,006 | 3,248,968 | 3,049,332 | 298,637 |

* The *N* of journals is 20,554 for the period 1996-2012

** corrected for single citation links.

JCR data were harvested from the two JCR files: one for the SCI 2012 (covering 8,471 journals) and the other for the SSCI 2012 (3,047 journals; see the right-side columns of Table 1). The two files were first merged so that the 582 journals which are covered by both the SCI and the SSCI are counted only once. The category “All others” is considered as a missing value. Note that the resulting file contains values similar to the Scopus data (after correction for the single occurrences): 37.8 million citations in WoS *versus* 36.8 million in Scopus. Since the number of journals in Scopus is almost twice as high, one would expect the density in this network to be much lower.

The method of calculating journal-to-journal-citation matrices based on Scopus data is essentially different from the one applied to WoS data. For example, if a journal is founded in 2000, and indexed in both databases as from publication year 2005, JCR 2005 would count citations to all five previous years, but Scopus count would only include citations to the 2005 annual volume. The generation of a journal-journal matrix by SCImago using Scopus data is based on citation links among documents included in the Scopus database, whereas JCR includes non-indexed journals using the abbreviated journal names in the cited references (Moed & Van Leeuwen, 1995 and 1996). We return to the consequences of this difference in the respective procedures in a later section.

[4] “Invalid citation relations” are relations where the sequence number of the cited journal is not referring to a journal in the master list of journals. These mismatches are incidental.

[5] JCR 2012 data contain only 189 single-citation relations. However, the number of relations with the value of 2 is 709,426 so that one can assume that single-citation relations are incidental. Single citation relations are usually subsumed under the category “All others” when the analysis is pursued from the “citing” side.

## 3. Methods

The data were organized under a system for relational database management and then exported in the format of Pajek.[6] Pajek can be used for the analysis and visualization of large networks (De Nooy *et al.*, 2011). Pajek (v3.11) is fully integrated with VOSviewer (v1.5.5). Note that two files are exported for each database: one with the original (count) data for the network analysis and another with the cosine-normalized data for the visualizations.

### *3.1 Maps and visualizations*

For the visualizations, we elaborate on the previous maps that were made available online for JCR 2011 (at http://www.leydesdorff.net/journals11; Leydesdorff, Rafols, & Chen, 2013) and Scopus 1996-2012 (at http://www.leydesdorff.net/scopus_ovl ; Leydesdorff, de Moya-Anegón & Guerrero-Bote, in press), respectively. The new routines are kept methodologically similar to the previous ones so that one can also compare results diachronically. The maps for 2012 data are available for interactive usage at http://www.leydesdorff.net/journals12 for WoS and http://www.leydesdorff.net/scopus12 for Scopus data. In addition to webstarting the complete maps (see below), the user can overlay downloads from either database (Scopus or WoS) and generate overlays interactively. For more technical details the reader is referred to the websites or the Appendices.

[6] Pajek is a program for network analysis and visualization freely available for non-commercial us at http://pajek.imfm.si/doku.php?id=download .

In the case of WoS data, we use—as previously—cosine > 0.2 as a threshold because this reduces the file size in Pajek from 623 to 39 Mbyte, while it leaves us with a giant component of 10,549 of the 10,936 journals—that is, 96.5% of the journals. This smaller-sized file could be imported into VOSviewer for mapping and clustering using 8 GB on a laptop under a 64-bits operating system (Windows 7). The much larger number of journals in the case of Scopus data made us use (as before) a Unix-based environment with 24 GB at the Amsterdam Computer Center SARA. Since there were no longer limitations to the size in this latter environment, no threshold is necessary. The largest component which is analyzed and mapped using VOSviewer under Unix contains 18,160 (90.0%) of the 20,172 journals in the database (Table 2). We further removed three journals in the case of WoS data and six in the case of Scopus data because they distorted the visualization in VOSviewer as outliers.

**Table 2**: Statistics used for the visualization in VOSviewer

| | *JCR-WoS 2012* | *Scopus 2012* |
|---|---|---|
| Giant component | 10,545 | 18,160 |
| After correction for visual outliers | 10,542 | 18,154 |
| *N* of clusters (Blondel *et al.*, 2008) | 12 | 65 |
| *N* of clusters (VOSviewer) | 11 | 47 |
| Modularity *Q* | 0.569 | 0.694 |

As noted, the results of this analysis are strictly analogous to the ones for the aggregated set of Scopus 1996-2012 data published by Leydesdorff *et al.* (in press), and similarly analogous to the map based on JCR 2011 used by Leydesdorff *et al.* (2013). However, the two maps for 2012 (and the underlying matrices) can also be compared because they are both representations of otherwise similar citation data for the same year. For methodological reasons, we also generated the maps without a cosine threshold in the case of JCR data and with the threshold of cosine > 0.2 in the case of Scopus data. However, we postpone the discussion of these variants to the discussion section.

VOSviewer was used for these visualizations for the following reasons (van Eck & Waltman, 2010):

- The continuity with previous visualizations of similar data for other years; we had routines available that were adapted for processing this new data.
- The problem of the cluttering of a large number of labels in the visualization is solved in VOSviewer by fading the less-important nodes gradually. However, all nodes can be foregrounded by clicking on them. One can also solve the problem of potential cluttering in the case of a large number of labels by setting node labels to proportionally-sized fonts in Gephi,[7] but this may still lead to difficulties in the visual reading of the map (Leydesdorff *et al.*, 2011, at pp. 2420f.).
- As noted, VOSviewer is now seamlessly interfaced with Pajek in both the directions for import and export of data. We use Pajek for the network analysis and VOSviewer for the visualization.
- The resulting maps can be exported locally or also uploaded and webstarted. We used the clustering algorithm of VOSviewer as a default, but this can be changed by the user as explained in Leydesdorff *et al.* (2013, Appendix 2, p. 2586). Both files contain also the results of the community-finding based on Blondel *et al.*'s (2008) algorithm in Pajek (as an additional field named "Blondel").

[7] Gephi is a network visualization program freely available at https://gephi.org/ .

*3.2 Network analysis*

For the network analysis of the two matrices, we distinguish between the journals that are common to the two databases, and the journals unique to each of them. The matching between the two databases was done stepwise:

1. After correction for duplicate ISSN numbers, journals with identical ISSN numbers were identified as identical (cf. Gavel & Iselid, 2008). Thus, 10,276 journals could be matched;
2. For the unmatched journals, journals with identical titles in the two databases were considered to be identical; this applied to 196 journals;
3. Fuzzy-string matching with the Ratcliff/Obershelp algorithm (Ratcliff & Metzener, 1988) was used on the journals remaining after the second step, using the larger list of Scopus journals as input to the matching.[8] Journals with similar titles but different ISSN numbers were matched only if the ISSN numbers were found to refer to the same journal in different formats, such as print versus an electronic version. Thus another 52 journals could be matched.

In total 10,524 journals could thus be matched. During the process of matching WoS journals to Scopus journals, we identified another six duplicate journals in WoS[9] and one duplicate journal in our Scopus data, so that the number of journals used in the comparison is slightly lower than reported Table 1, namely 10,930 journals in WoS and 20,553 journals in Scopus. The unique ones are 406 journals (3.7%) in the case of WoS and 10,029 (48.8%) in the case of Scopus. In other words, almost all journals in WoS are also covered by Scopus. Additionally, it can be notified that 326 (of the 10,936) journals in WoS were not processed in terms of citations in JCR 2012 from the citing side, but only included when among the cited references. Of these 326 journals, 88 are not covered by Scopus.

[8] The algorithm is implemented in Access by Tom van Stiphout at http://accessmvp.com/tomvanstiphout/simil.htm .
[9] These six journals are included with their previous name in languages other than English, but no longer processed in their citing patterns. For example, *Biomedizinische Technik* is now covered as *Biomedical Engineering-Biomedizinische Technik*. These journals continued using the same ISSN.

We performed the network analysis for both the shared sets and the unique sets in either database, and compared the resulting networks in terms of parameters such as density, the numbers of (weak)[10] network components, isolates, sizes of largest components, average indegree, and indegree centralization. Because both the numbers of links and their weights in terms of numbers of citations are available, this data allows us to rank the journals both in terms of indegree centrality (that is, number of journals citing a journal) and by the numbers of citations, and thus to show the relative positions of shared and unique journals in each database. The differences in the ranks among shared journals will further be analyzed.

**4. Results**

As in the methods section (above), we discuss the results first of the visualizations and mapping, and then of the network analysis. In a final section, we will visualize the shared and unique journals in Scopus as two overlays to the base map for all Scopus journals. Rao-Stirling diversity values of the two sets provide us with a statistic to express the variety and disparity in terms of the coverage across the maps (Leydesdorff *et al.*, 2013:2578; cf. Rafols & Meyer, 2010; Rao, 1982; Stirling, 2007).

[10] Unlike a strong component, a weak component does not require that the citation relation be reciprocal between two journals.

### *4.1 Global maps*

Figure 1 shows the base map for the 10,542 journals (96.4%) included in the largest component of JCR 2012 (see Table 2). Its shape and coverage is very similar to the map for 2011 (Leydesdorff *et al.*, 2013, Fig. 1 at p. 2575).[11] Although 11 clusters are distinguished (as against 12 in 2011), the overall pattern of the partitioning (and hence the coloring) is also very similar. This reproduction of a base map in two different years—using the same methods—generates confidence in the validity of the technique and the reliability of this data.

[11] The corresponding map for 2011 can be viewed directly in VOSviewer via webStart at http://www.vosviewer.com/vosviewer.php?map=http://www.leydesdorff.net/journals11/citing_all.txt&label_size=1.0&label_size_variation=0.3 .

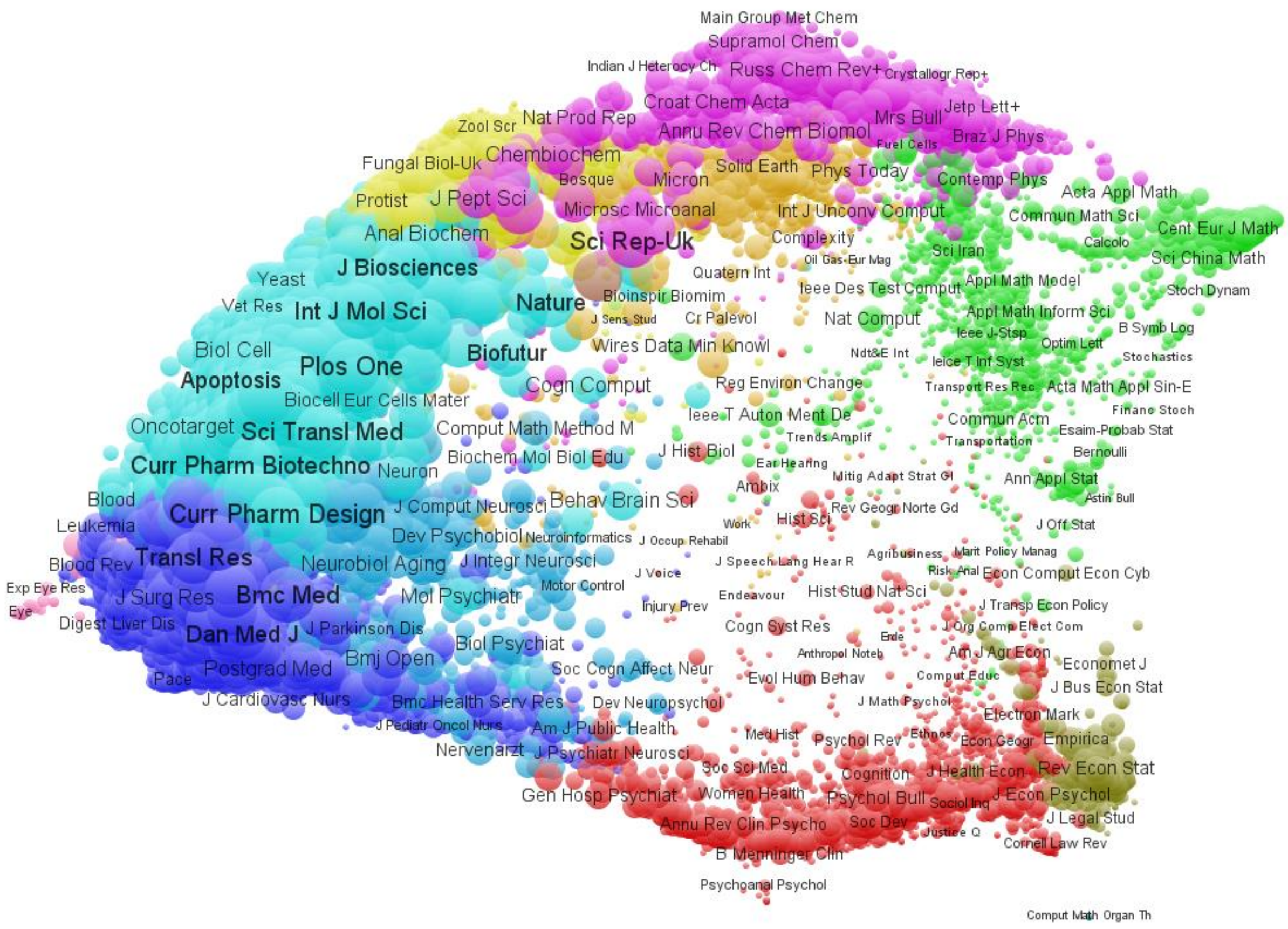

**Figure 1**: Citing patterns of 10,542 journals in JCR 2012 visualized as a base map; cosine > .2; colors correspond to 11 communities distinguished by VOSviewer; available for webstart at http://www.vosviewer.com/vosviewer.php?map=http://www.leydesdorff.net/journals12/jcr12.txt

The map based on Scopus data 2012 (Figure 2) is also not so different from the previously published map based on aggregated Scopus data 1996-2012 (Leydesdorff *et al.*, in press: Figure 3).[12] As expected, the number of communities detected (using VOSviewer) in 2012 is larger (42)[13] than the 27 communities detected in the aggregate of 1996 to 2012 because of the larger density in the latter map. Compared to the aggregated map, the base map for 2012 was deliberately flipped (by us) along the vertical axis because thus the similarity with the JCR-based map (in Figure 1) is more easily recognizable. The tail of the humanities journals at the bottom right is absent from the JCR-based maps, while the A&HCI is not included in JCR.

---

[12] The base map of aggregated citation relations among 19,600 journals included in Scopus 1996-2012 is available at http://www.vosviewer.com/vosviewer.php?map=http://www.leydesdorff.net/scopus_ovl/basemap.txt

[13] Table 2 mentions 47 clusters, but this is the number before the removal of six outliers.

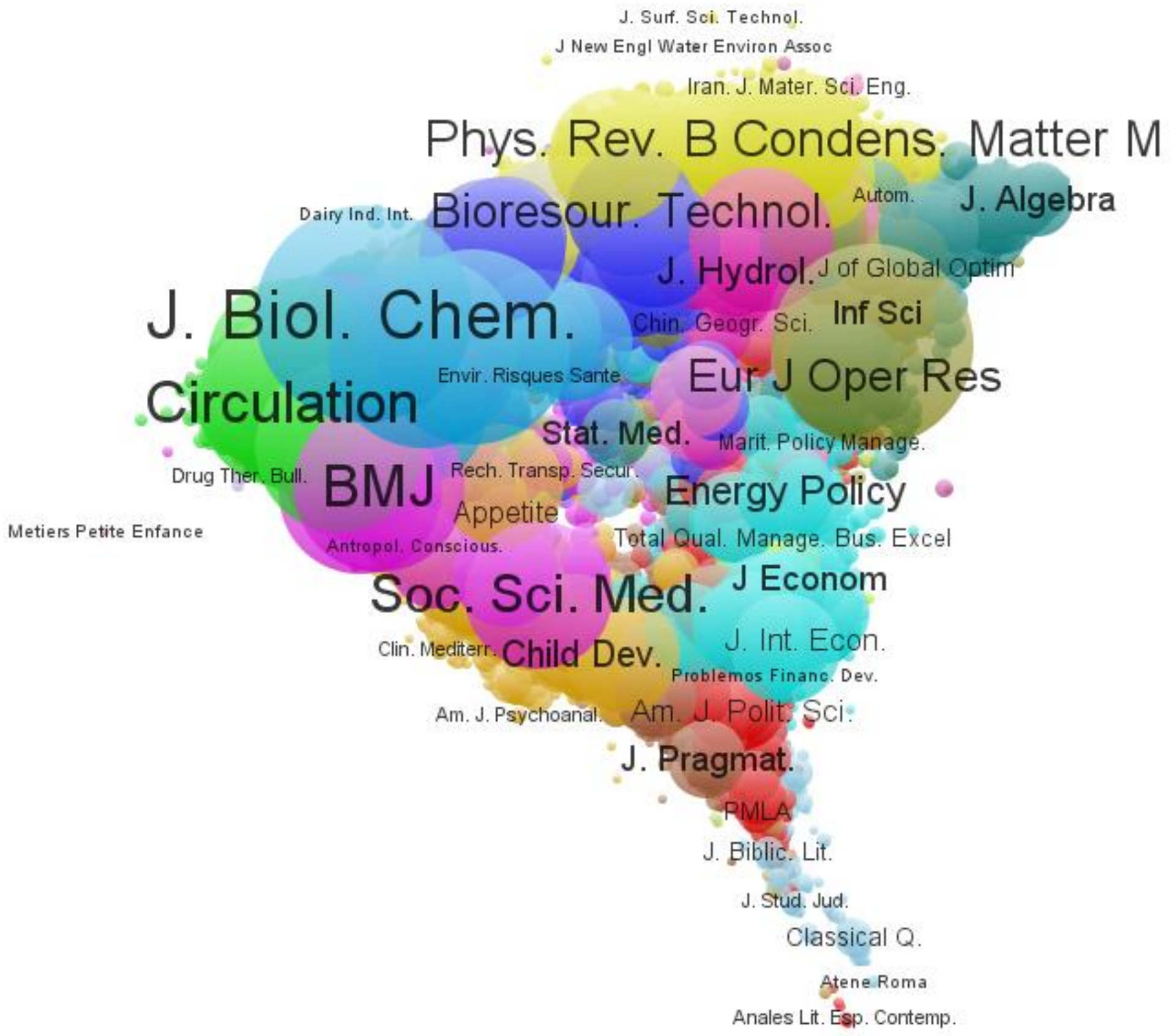

**Figure 2**: Citing patterns of 18,154 journals in Scopus 2012 visualized as a base map; colors correspond to 42 communities distinguished by VOSviewer; available for webstart at http://www.vosviewer.com/vosviewer.php?map=http://www.leydesdorff.net/scopus12/scopus12.txt.

The similarity between the maps can be made more visible by rotating the map 20 degrees anti-clockwise and normalizing the weights of the nodes differently, as in Figure 3.

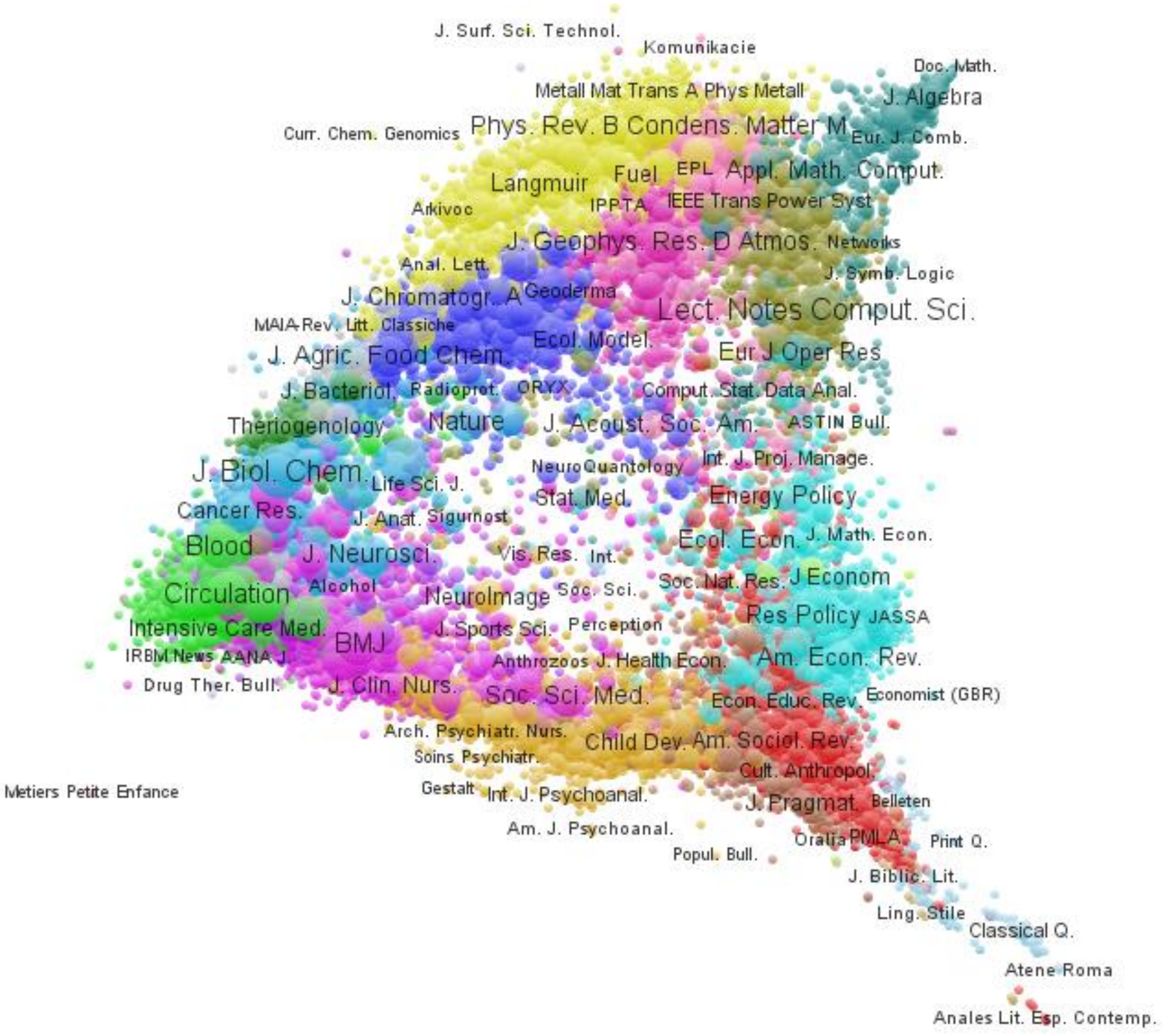

**Figure 3**: Citing patterns of 18,154 journals in Scopus 2012; colors correspond to 42 communities distinguished by VOSviewer; weights normalized and rotated -20 degrees when compared to Figure 2.

The largest component in this cosine-normalized network of 20,172 journals in 2012 contains 18,154 journals (90.0%). Although this coverage is smaller than the coverage in JCR data (96.4%), it is still unexpectedly high given the huge enlargement of the journal set in Scopus when compared with JCR. However, we did not apply a threshold to the cosine in the case of Scopus data (after discarding the single citation relations), and thus very small cosine values potentially based on incidental citation relations (above 1) were also included when generating this map. This may have caused the relative dominance of the large journals in Figure 2.

However, one can change this effect in the visualization by turning the “label-size variation” off in VOSviewer interactively or as follows: http://www.vosviewer.com/vosviewer.php?map=http://www.leydesdorff.net/scopus12/scopus12.txt&label_size_variation=0 . In the case of overlays (see below), the node sizes are anyhow determined differently by the logarithms of the numbers of publications in the sample(s) under study.

### *4.2 Comparing overlays between JCR and Scopus*

The base maps can be used to position sets of documents (e.g., a portfolio) in terms of their disciplinary composition. In the two previous studies, we used datasets generated by Rafols *et al.* (2012) for the study of interdisciplinarity in which the Science and Technology Policy Research Unit (SPRU) at the University of Sussex was compared with the London Business School (LBS). These two sets were also used by Leydesdorff *et al.* (2013) to show the respective portfolios projected on the base map of WoS (Figures 3a and 3b in that study), and in Leydesdorff *et al.* (in press) for comparing the Scopus 1996-2012 set with JCR-based maps for 2011 (Figures 8a and 8b of that study). In order to maximize the possibilities for comparisons *ceteris paribus*, we use these same sets of documents also as examples in this study.

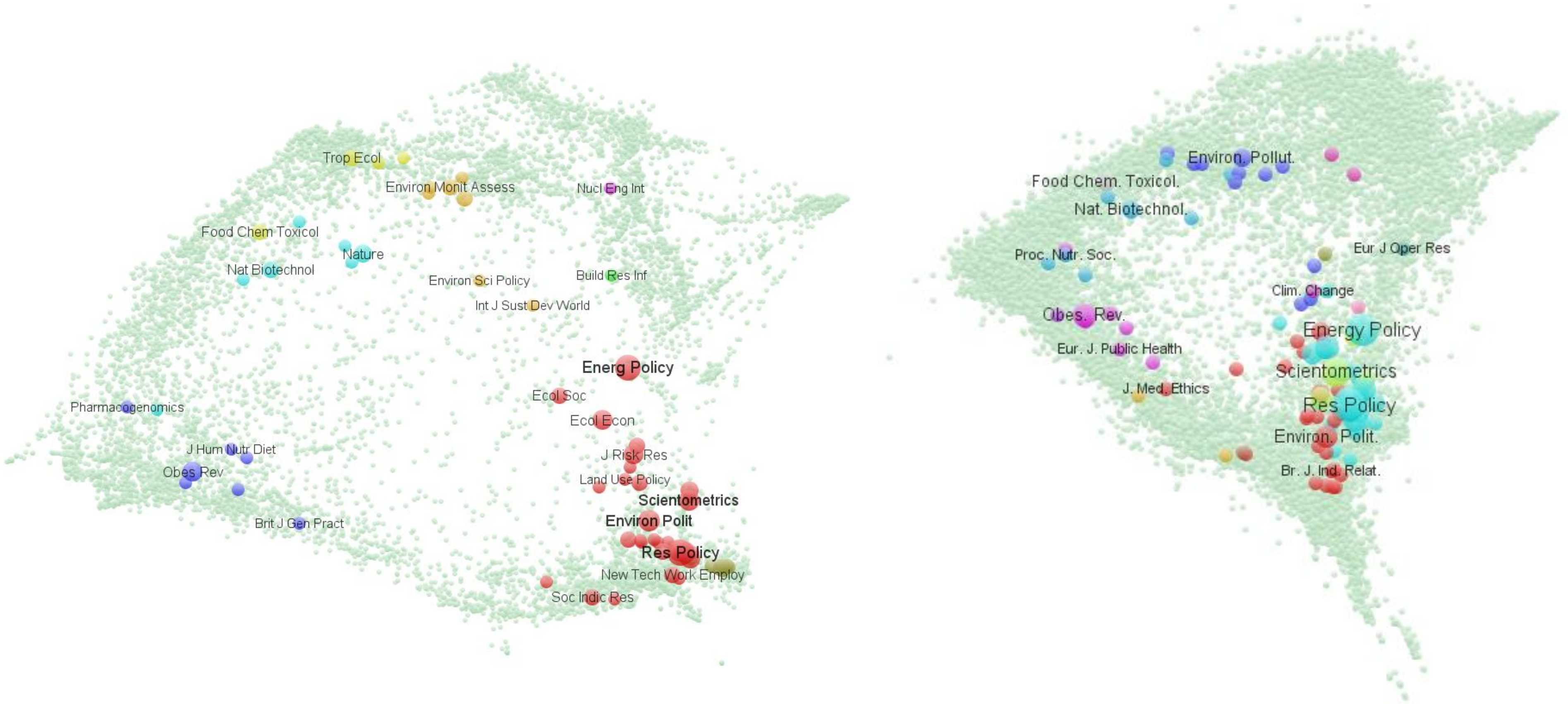

**Figures 4a and b**:

a) WoS-based overlay map 2012 of journal publication portfolios from 2006 to 2010 of the Science and Technology Policy Research Unit SPRU at the University of Sussex (on the left; *N* = 155; available at http://www.vosviewer.com/vosviewer.php?map=http://www.leydesdorff.net/journals12/spru.txt&label_size=1.35 )

a) Scopus-based overlay map 2012 of journal publication portfolios from 2006 to 2010 of the Science and Technology Policy Research Unit SPRU at the University of Sussex (on the left; *N* = 268; available at http://www.vosviewer.com/vosviewer.php?map=http://www.leydesdorff.net/scopus12/spru.txt&label_size=1.35 )

Figures 4a and b show the sets downloaded from WoS (*N* = 155) and Scopus (*N* =268), respectively, for the Science and Technology Policy Research Unit SPRU of the University of Sussex. The sets of the LBS downloaded from WoS (*N* = 384) and Scopus (*N* = 715) provide similar options for the comparison—not shown here, but available for webstarting at http://www.vosviewer.com/vosviewer.php?map=http://www.leydesdorff.net/journals12/lbs.txt&label_size=1.35 and http://www.vosviewer.com/vosviewer.php?map=http://www.leydesdorff.net/scopus12/lbs.txt&label_size=1.35 , respectively.

**Table 3**: Rao-Stirling diversity for SPRU and LBS documents (2006-2010) in both the 2011 and 2012 maps based on annual JCR data, and the two Scopus maps.

| | JCR 2012 (a) | JCR 2011 (b) | *N* *WoS* | Scopus 2012 (c) | Scopus 1996-2012 (d) | *N* *Scopus* |
|---|---|---|---|---|---|---|
| *SPRU* | 0.2134 | 0.2175 | 155 | 0.1219 | 0.1489 | 268 |
| *LBS* | 0.0893 | 0.0922 | 348 | 0.0863 | 0.0917 | 715 |

Using Rao-Stirling diversity values (that are generated automatically using the corresponding routines; see the Appendix), Table 3 shows that the values of this indicator are virtually the same for JCR 2011 and JCR 2012. As noted before, this map seems robust. In the case of the Scopus database, however, we do not have a corresponding comparison, but the values using exclusively the subset of 2012 data are systematically a bit lower than the values using the aggregate of 1996-2012. Since both overlays are based on the same data in the downloads, the distribution across the journals is the same, and the smaller values are due only to a decrease in relative distances when we consider only a single year compared to a broader set.

Although the values of Rao-Stirling diversity are almost identical for LBS, this value is much higher for SPRU in JCR (Table 3). Thus, the inclusion of a larger set of journals in the case of Scopus leads to a more disciplinarily oriented representation of this research unit (SPRU). In other words, SPRU staff publishes also in disciplinarily core journals that are not included in WoS, but are included in Scopus.

### *4.3 Network analysis*

Using the 10,524 journals that could be matched between WoS and Scopus, one can construct the citation networks of the set of journals shared between the two databases and compare their structures. Furthermore, we are able to analyze the structures of the journal sets that are unique to one of the two databases. The results show some consequences of the database choices for the structure of the citation network, and can therefore be relevant for indicators based on their respective structures.

**Table 4**: Comparison of WoS and Scopus citation networks

| | | WoS | | | Scopus | |
|---|---|---|---|---|---|---|
| | Shared | Unique | Overall | Shared | Unique | Overall |
| Journals *Number* | 10,524 | 406 | 10,930 | 10,524 | 10,029 | 20,553 |
| *% of overall network* | 96,3% | 3.7% | 100% | 51.2% | 48.8% | 100% |
| Average number of citations received (per journal) | 3,511.0 | 1064.8 | 3,454.7 | 2,814.2 | 60.5 | 1,683.2 |
| *Structures within (sub)networks* | | | | | | |
| Density* | .020 | .004 | .020 | .018 | .001 | .006 |
| Average indegree* | 215.6 | 1.9 | 214.1 | 190.1 | 5.5 | 130.1 |
| Indegree centralization* | .58 | .07 | .57 | .59 | .06 | .41 |
| Number of weak components | 16 | 167 | 15 | 43 | 2,122 | 1,118 |
| Size of largest weak component (% of Journals) | 99.9% | 51.5% | 99.9% | 99.6% | 78.5% | 94.5% |
| Number of isolates | 15 | 150 | 14 | 42 | 2,091 | 1,113 |

* Within-journal self-citations (network loops or the main diagonal) are excluded from the network analysis.

Table 4 summarizes the characteristics of the networks of shared and unique journals for the two databases.

### 4.3.1 The 10,524 shared journals

Let us first compare the networks among the journals that are included in both databases (the columns labeled 'Shared' in Table 3). The two databases are mainly different in terms of their density: the WoS network of shared journals is 10% denser (.020 *versus* .018 in Scopus) and, as a consequence, journals receive on average citations from more journals in the JCR set (the average indegree is 215.6 for WoS and 190.1 for Scopus). WoS registers almost 20% more citations among the same set of journals (on average 3,511.0 per journal) than Scopus (on average 2,814.2 per journal). This difference results in more citation linkages among journals, fewer isolates (journals not citing or being cited by other journals in the set), and a more connected network (a relatively larger main component) in WoS when compared with Scopus.

This rather large difference in terms of registered citations can be explained in terms of so-called "active" citations (Zitt & Small, 2008). Whereas WoS *registers* all references equally (Moed & van Leeuwen, 1995 and 1996), Scopus *identifies* references in terms of the cited documents. However, publications from before 1996 are not identified in Scopus. The tail of historical references before 1996 is therefore not processed when aggregated into the journal-journal citation matrix.[17] At the occasion of the celebration of ten years of Scopus (March 27, 2014), the

[17] JCR data allow us to compute the percentage of citations for each of the last ten years (2003-2012) and the remaining tail of references to publications before 2003. This latter set is 13,678,737 citations or 36.7% of

Scopus team most recently announced the launch of the Scopus Cited References Expansion project: "After extensive evaluation of feedback from the research community, internal discussion and operational documentation, our content team successfully made the investment case to include cited references in the Scopus database – going back to 1970 for pre-1996 content!" (retrieved from http://blog.scopus.com/posts/scopus-to-add-cited-references-for-pre-1996-content on May 15, 2014).[18]

The two networks of shared journals have comparable centralization, that is, the indegree variation is nearly equal (.58 for WOS and .59 for Scopus). Assuming that indegree is an indicator of journal standing—being cited by more journals reflects higher standing—similar indegree centralization scores show that the variation in standing is more or less equal for shared journals in both databases. A high rank correlation between journal indegree in WoS and Scopus—Spearman's rho = .974 for indegree and .975 for total citations—also indicates that the choice of the database does not matter much to the overall ranking of journals. These overall correlations, however, do not preclude large rank differences for individual journals (Figure 5 and Table 5).

---

37,295,253 in the shared set of 10,524 journals.

[18] See also: *Against the Grain*, at http://www.against-the-grain.com/2014/03/newsflash-celebrating-10-years-scopus-announces-major-archival-expansion-program/ and a press release at http://www.elsevier.com/about/press-releases/science-and-technology/elsevier-announces-launch-of-program-to-include-cited-references-for-archival-content-in-scopus .

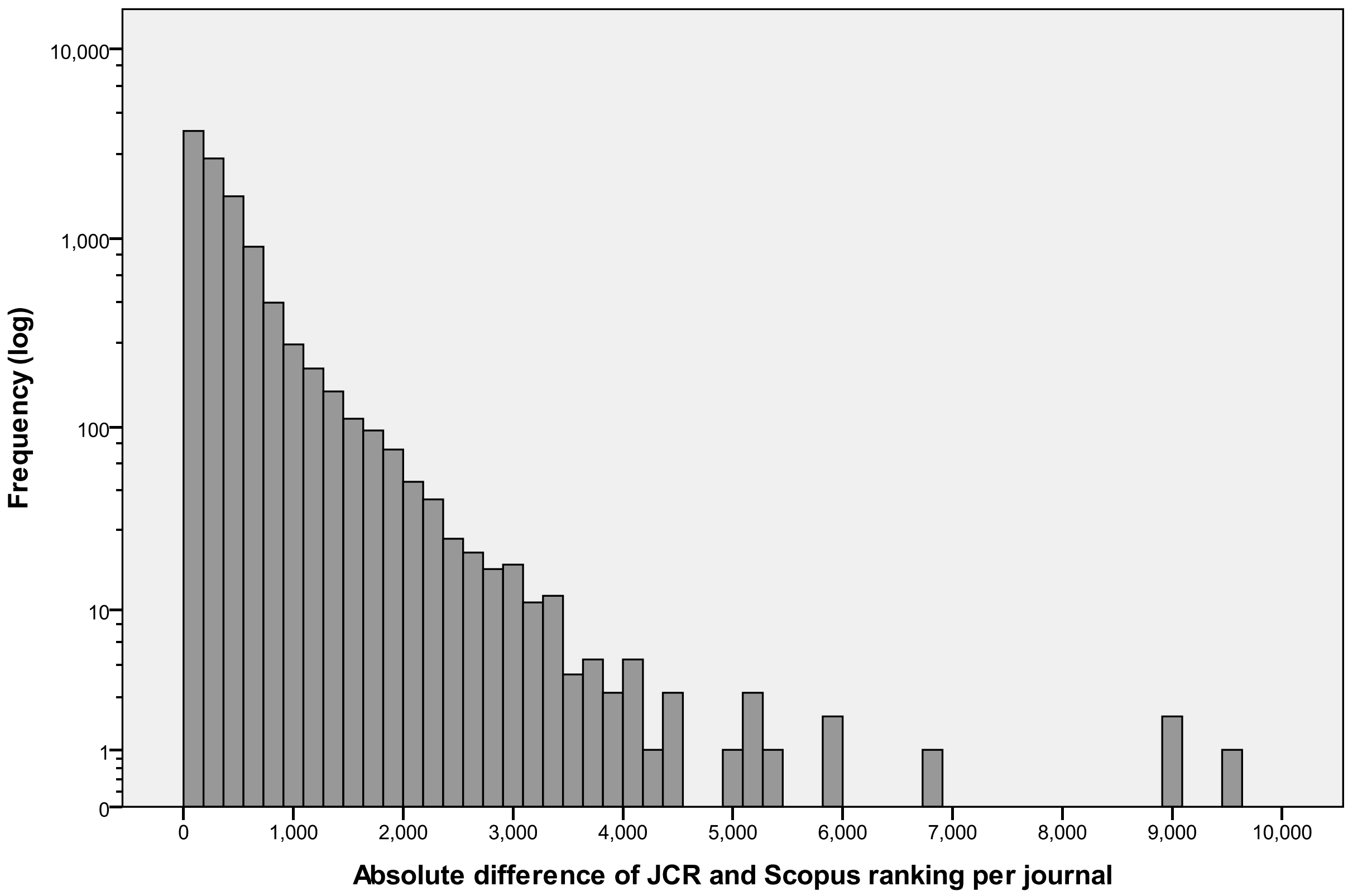

**Figure 5**: Rank differences in terms of indegree for 10,524 journals shared among WoS and Scopus.

Figure 5 shows the absolute differences in rankings between the indegree in JCR and Scopus on a logarithmic scale. (The corresponding figure for a ranking based on total citations is very similar.) Although thousands of journals have relatively small differences with respect to their indegree ranks in the two databases (to the left in the graph), dozens of journals have ranks that differ by at least two thousand positions, and a handful of journals occupy positions at the top in one database (WoS) but at the bottom of the rank list in the other database (Scopus) with rank differences of greater than 5,000. In Table 5, the top-20 journals in terms of these differences in ranks are listed. Of course, we are not able to exclude that some of these journals may have been erroneously matched, and the noted effect of cutting of the historical tails of the distributions in Scopus before 1996 may also play a role here.

**Table 5**: Top 20 journals in terms of rank difference between JCR and Scopus in terms of citations (self-citations included; on the left side) and indegrees (on the right side).

| *Journal abbreviations (according to Scopus)* | *Citations JCR* | *Citations Scopus* | *Rank difference* | *Journal abbreviations (according to JCR)* | *Indegree JCR* | *Indegree Scopus* | *Rank difference* |
|---|---|---|---|---|---|---|---|
| Mutat. Res. | 7676 | 6 | 9212 | Mutat Res-Fund Mol M | 967 | 2 | 9619 |
| Oral Surg. Oral Med. Oral | 10484 | 23 | 8963 | Or Surg Or Med Or Pa | 634 | 3 | 9025 |
| World Health Organ Tech R | 2200 | 2 | 7691 | Who Tech Rep Ser | 496 | 1 | 8977 |
| Solid. State Phys. Adv. R | 1630 | 43 | 6049 | Dtsch Arztebl Int | 174 | 0 | 6768 |
| Lect. Notes Math. | 8352 | 233 | 5913 | Solid State Phys | 237 | 14 | 5943 |
| Dtsch Arztebl Int | 872 | 0 | 5907 | Ieee T Comp Pack Man | 271 | 20 | 5825 |
| IEEE Trans. Compon. Packa | 2839 | 117 | 5793 | Cell Oncol | 112 | 1 | 5318 |
| Trans. ASABE | 5962 | 237 | 5565 | Colloq Math-Warsaw | 126 | 4 | 5187 |
| Geoscientific Model Dev. | 579 | 6 | 4733 | Collegium Antropol | 26 | 170 | 5132 |
| Eur. J. Wood Wood Prod. | 164 | 1079 | 4545 | T Asabe | 446 | 51 | 5115 |
| Earth Environ. Sci. Trans | 1066 | 78 | 4481 | Lect Notes Math | 590 | 61 | 5079 |
| Paleontogr. Abt. B Palaeo | 494 | 4 | 4448 | Arch Environ Occup H | 250 | 37 | 4510 |
| Colloq. Math. | 600 | 18 | 4429 | Evol Bioinform | 201 | 29 | 4438 |
| Evol. Bioinformatics | 1266 | 110 | 4390 | Nonlinear Oscil | 88 | 3 | 4387 |
| Archiv. Environ. Occup. H | 1386 | 146 | 4143 | Sb Math+ | 251 | 43 | 4200 |
| Cell Oncol (Dordr) | 414 | 2 | 4139 | Nation | 76 | 2 | 4082 |
| Funct. Anal. Appl. | 1434 | 164 | 4001 | St Petersb Math J+ | 76 | 2 | 4082 |
| Eur Space Agency Bull | 525 | 23 | 3999 | Nebr Sym Motiv | 112 | 12 | 4061 |
| Nonlinear Oscill. | 487 | 17 | 3980 | Funct Anal Appl+ | 234 | 43 | 4033 |
| Sb. Math | 1840 | 234 | 3836 | Brief Funct Genomics | 172 | 29 | 4016 |

Note 1: The abbreviations were truncated after 20 characters in JCR and after 25 in Scopus.
Note 2: In the journal abbreviations according to JCR, the “+” flags that the journal is published in the Cyrillic alphabet (that is, in Russian).

*Collegium Antropologicum* is the only journal with a higher indegree in Scopus than in JCR, and the *European Journal of Wood and Wood Products* was cited more often in the Scopus domain than in JCR. Probably these journals were cited in the arts and humanities journals that are included in Scopus but not in JCR. Alternatively, a journal may have been included in JCR more recently than in Scopus or vice versa.

### 4.3.2 Unique journals

In contrast to the networks among shared journals, the networks of citations among unique journals are very sparse (see the columns headed 'Unique' in Table 3). There are few citations and citation linkages among journals that are included in the one database and not in the other. For example, the average indegree of unique journals within these sets is only 1.9 in WOS and 5.5 in Scopus (Table 4). This is by an order of magnitude smaller than the average indegree in the set of shared journals. Consequently, the subnetworks of unique journals are much less connected than the subnetworks of shared journals; they contain more components and isolated journals; and smaller main components. These results apply to both databases, suggesting that the shared journals constitute the core of the citation network and the database-unique journals can be considered as belonging to the periphery.

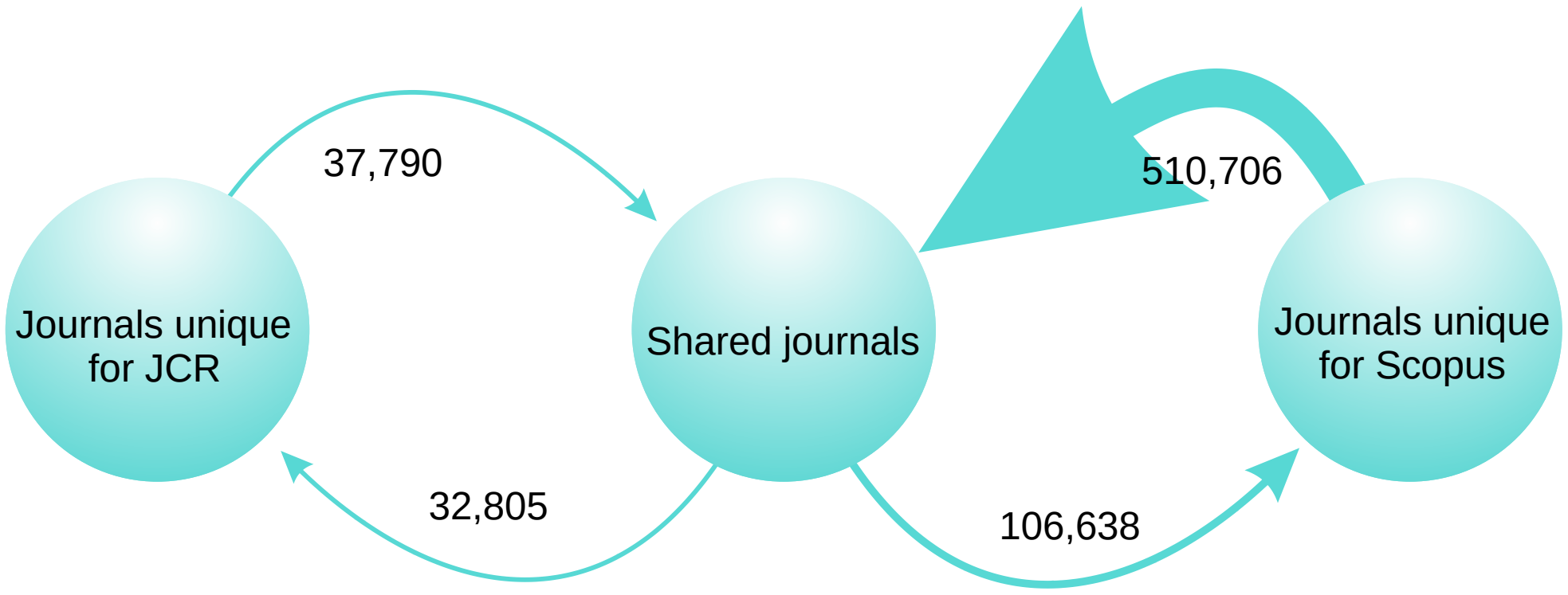

**Figure 6**: Citation relations among shared and unique journals in JCR (left side) and Scopus (right side).

In the case of Scopus, this suggestion is confirmed by the direction of citations relations as shown in Figure 6. The citations predominantly point from the unique Scopus journals to the shared journals. If receiving ("being cited") rather than sending citations signals higher status, the shared journals have higher status than the journals that are unique to Scopus. In contrast, the citation flows in both directions are more balanced in the case of WoS, so the 406 journals that are unique to WoS need not be less important than shared journals; they are just not covered by Scopus. We conclude that the journals unique to Scopus are relatively unimportant for the *structure* of the network and that JCR includes the most important journals from the larger set of journals covered by Scopus.

For the readers interested in further details, we bring an Excel file online at http://www.leydesdorff.net/journals12/all_journals.xlsx which lists (*i*) the 10,524 journals matched (by us!), (*ii*) the 406 journals unique to WoS, and (*iii*) the 10,029 journals unique to Scopus. In addition to journal names, journal abbreviations, and ISSN numbers are provided for the 20,959 journals in total.

### *4.4 The disciplinary composition of the unique journals of Scopus*

Our final question about the large set of unique Scopus journals is whether they concentrate in particular communities or specialties. Does Scopus cover certain scientific fields more extensively than JCR? To this end, we applied the base-map technique presented above to the sets of unique and shared journals in an overlay. Figure 7 shows that unique Scopus journals are found across all communities and specialties; the overlay offers an impression very similar to Figure 2.

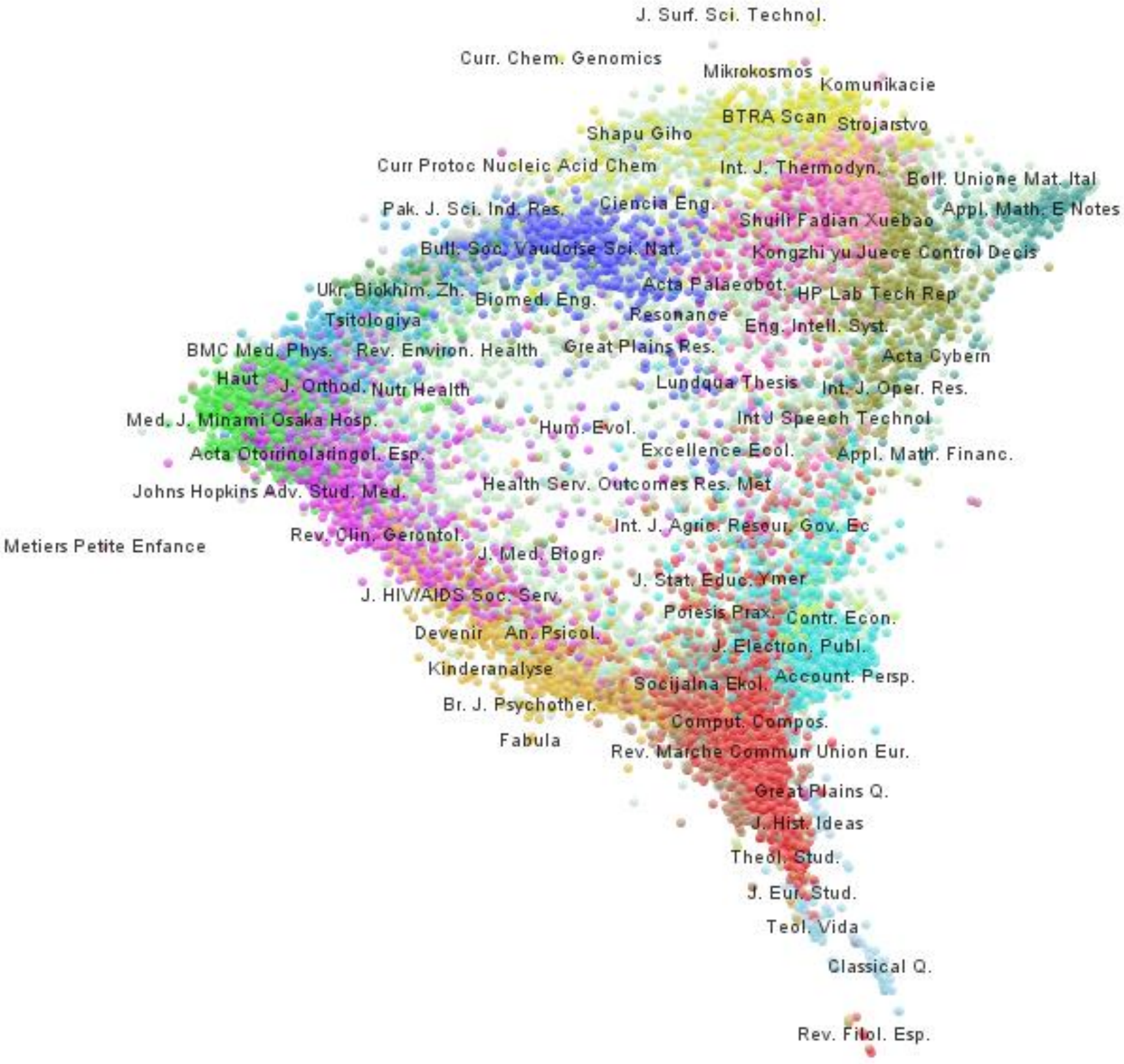

**Figure 7:** Overlay of 7,824 journals unique to Scopus 2012; Rao-Stirling diversity = 0.2679; available at

http://www.vosviewer.com/vosviewer.php?map=http://www.leydesdorff.net/scopus12/unique.txt&label_size=1.35

A comparative view can be obtained by depicting the density of journals as heat maps in Figures 8a and 8b, respectively. The density map for the journals that are shared with JCR (Figure 8a) is rather similar to the map for journals that are unique to Scopus (Figure 8b). However, journals in the humanities are absent in the former while they are present in the latter, namely as a tail at the bottom (right) of the map. In addition, journals in the social sciences are relatively frequent among unique Scopus journals in comparison to journals shared with WoS; there is more red at the bottom of the torus. In contrast, journals in agriculture and bio-medicine are relatively more prominent among the journals shared between WoS and Scopus. The spread of the journals across the map as measured by Rao-Stirling diversity is very similar: 0.2509 for the shared journals and 0.2679 for the unique ones. The unique journals are a bit more dispersed across the map (for example, because of the presence of the humanities journals).

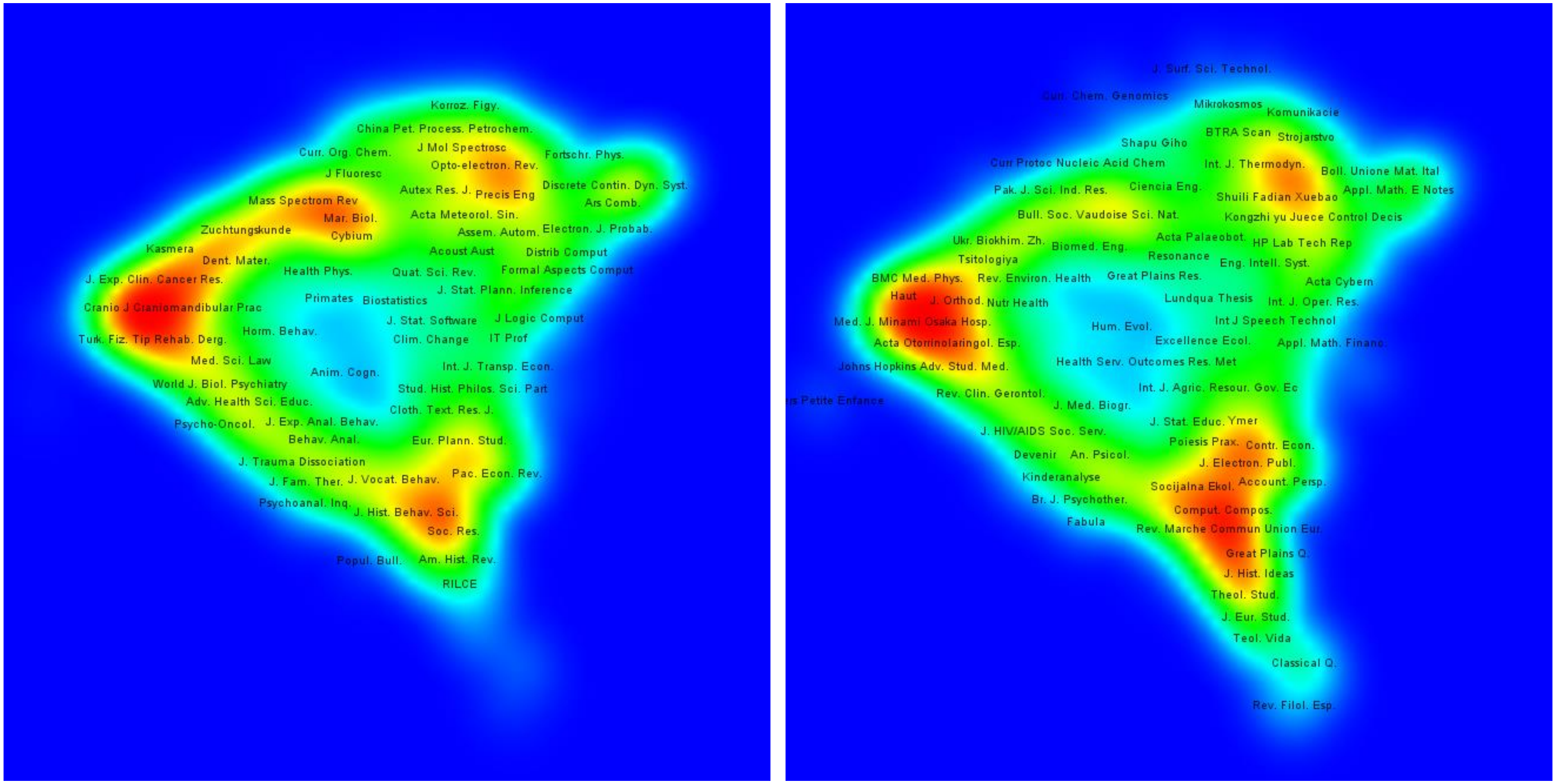

**Figures 8a and 8b**: Density maps of 10,330 journals in Scopus shared with JCR (left) *versus* 7,824 journals unique in Scopus (right). Rao-Stirling diversity is 0.2509 and 0.2679, respectively.

## 5. Conclusions

Of course, we knew that the set of journals covered in Scopus is much larger than in JCR and that therefore one would expect to find a lower density in Scopus. However, the (on average) 20% lower occupation of the citation matrix among the 10,000+ journals that are shared with JCR was unexpected. The explanation is that the "citable items" in Scopus and JCR are defined differently:

1. The references to non-source items are handled differently in Scopus and WoS because Scopus identifies source items as documents covered by the database. Since this database is only reliable after 1996, the historical tails of the citation distributions may be cut off differently depending on the partial coverage (or not) of some journals before 1996. The problem cannot easily be repaired because the identification of (cited) documents is part of the architecture of Scopus, and can in other respects be considered an advantage (Waltman & Van Eck, 2013). Anyhow, the Scimago group that provided us with the data does not process any data from before 1996 because these references cannot be identified in the Scopus database.
2. Some source documents such as obituaries are not covered by Scopus while these documents can be cited in WoS and then count as so-called excess citations to non-source items (Moed & Van Leeuwen, 1995 and 1996).

Leydesdorff, Hammerfelt & Salah (2011) argued that the citation field in the A&HCI of WoS was often so weak that one would be inclined to agree with the hesitation of Thomson-Reuters about generating a JCR for this database (cf. Garfield, 1982a and b; Lariviere *et al.,* 2006; Nederhof, 2006). Nevertheless, the base-maps and overlays are enriched with a tail for the arts and humanities in the case of Scopus when compared with JCR-based visualizations. The price for this extension is perhaps paid in terms of robustness. The replication of the base map for JCR 2011 and 2012 led to remarkably similar results both in terms of visualizations and analytics such as Rao-Stirling diversity. The JCR maps thus seem highly reliable to us, but the Scopus maps can provide richer visualizations. Thus, which one of the two databases to use in a project may depend on the research question (and contextual factors such as institutional access).

The argument of Garfield's (1971) "Law of Scattering" (cf. Bradford, 1934)—that the tails of the citation distributions of most journals are part of the core of other journals, and that therefore the journal structure is tightly knit—is confirmed by our analysis. The shared journals do the job more than the further extensions with unique journals. It is not incidental that 96.3% of the journals covered by JCR are also covered by Scopus. As noted, our matching may include some false positives or false negatives, but the percentage of overlap will remain high. A further extension of the journal horizon as in Google Scholar is, therefore, not needed for bibliometric reasons for "mapping science"; but one can argue for comprehensiveness, of course, for bibliographic reasons.

Given the importance of both Scopus and WoS for evaluation processes, the differences in ranks and thus in journal measures that are derived from these numbers (like impact factors, etc., in WoS or SNIP and SJR in Scopus) are worrisome although perhaps unavoidable. References in articles are riddled with typos and mistakes which may add up to considerable percentages of the total citations (Leydesdorff, 2008: Table 4 at p. 285). Despite the huge efforts of the producers of these databases to make corrections, the processing of this basic data remains error-prone. The specification of error and uncertainty is one of the tasks of statistical analysis, and therefore made us add an assessment using methods from social-network analysis to the two visualizations (for Scopus and WoS in 2012). The results are perhaps not so different from the intuition, but informative nevertheless. For example, the noted differences in the definitions of "cited items" between the two databases have implications for the respective journal indicators such as SNIP and SJR in Scopus, and JIF in WoS.

**Discussion**

The referees suggested that we controlled for the methodological effects of setting a threshold (cosine > 0.2) in the case of the JCR-based map versus using no threshold in the Scopus-based map. As noted, our choice for a threshold in the case of the JCR-based map was pragmatic because it reduced the size of the file by more than an order of magnitude while maintaining a large component. This latter condition is not fulfilled in the case of the Scopus data: the main component is reduced to 16,169 journals or 78.7% of the full set ($N$ = 20,554). The remaining journals are clustered by VOSviewer into 430 clusters. The resulting map is also graphically less attractive as a base-map than the one in Figure 4.

Using the full file of JCR data (without a threshold) provided us with a largest component of 10,610 journals among the 10,936 journals included in JCR (97.0%). This is a marginal improvement compared to the 96.5% (10,549) of the journals included in the main component above (see Table 2 and Figure 1). Note that the main component in this case covers virtually all journals that are processed since (as noted above) 325 journals were included when cited, but not processed citing. These journals therefore cannot be connected.

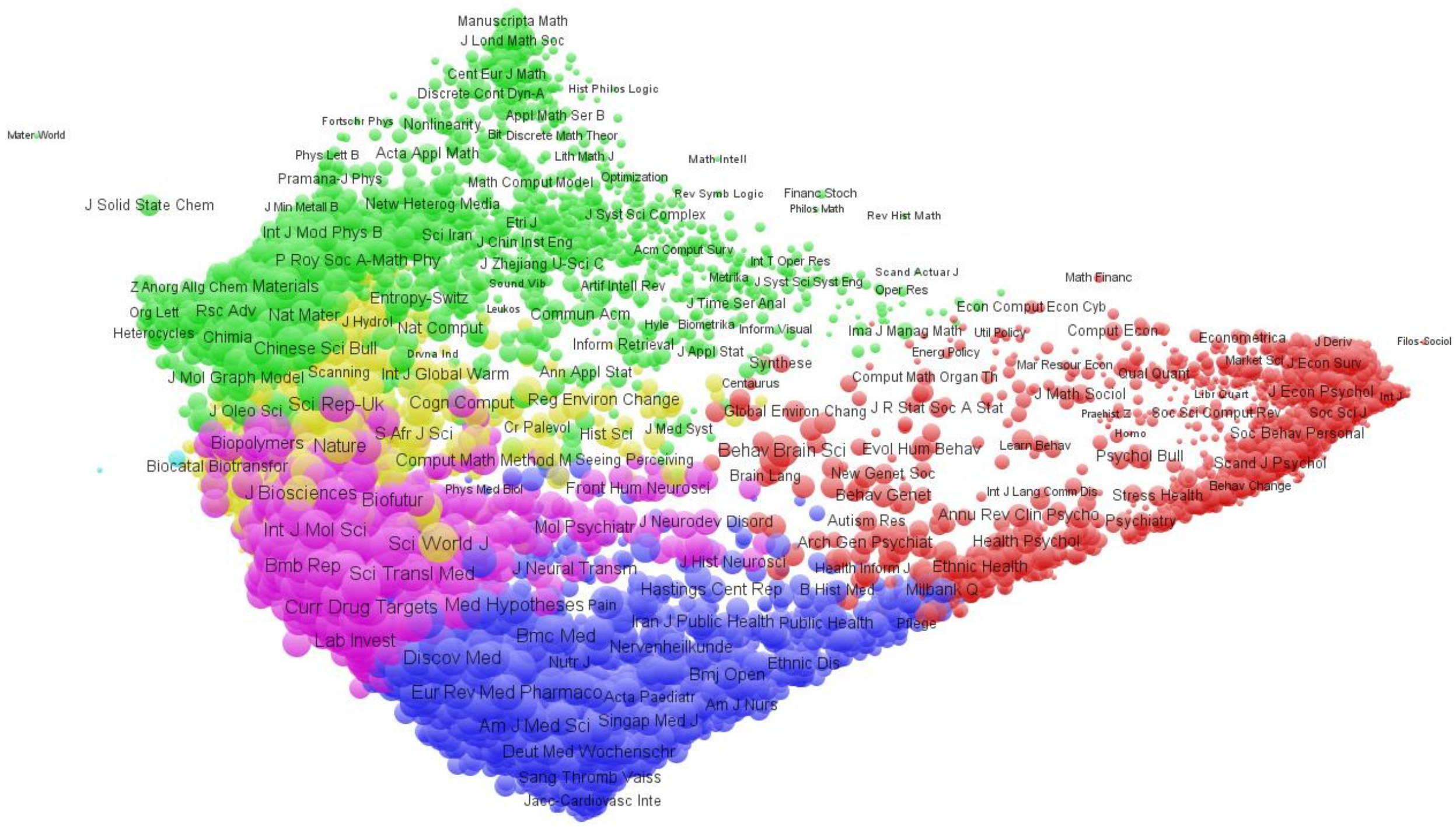

**Figure 9**: Citing patterns of 10,610 journals in JCR 2012; no threshold; VOSviewer used for the clustering and visualization; available at [http://www.vosviewer.com/vosviewer.php?map=http://www.leydesdorff.net/journals12/nothresh.txt](http://www.vosviewer.com/vosviewer.php?map=http://www.leydesdorff.net/journals12/nothresh.txt)

Using VOSviewer for the clustering and consequentially coloring in Figure 11, six major groups are distinguished of which one contains only two (biotechnology) journals. These communities are the major disciplinary groups such as: the social sciences and the humanities on the right side in red; the natural sciences and engineering at the top (green); the clinical sciences at the bottom (blue); the bio-medical sciences above that to the left (pink); and a more interdisciplinary group between the natural sciences and the bio-medical ones (yellow). As expected, the packing of this representation is denser than in Figure 1 because more cosine values are provided in the input file. However, one could use this map as a base-map if so wished.

In summary, we happen to have made fortunate parameter choices above when generating the base maps for JCR and Scopus data in Figures 1 and 2, respectively. With hindsight, the differently normalized and rotated representation in Figure 3 could have been a better option than Figure 2 itself. We worked with Figure 2 as the base map for Scopus data because we wished to keep the possibility to compare with the maps presented in Leydesdorff *et al.* (2014). However, other representations (e.g., using Gephi) remain equally possible.

**Acknowledgement**

We are grateful to Lykle Voort of the Amsterdam computer center SARA for his support, and grateful to Wim Meester of the Scopus team and two anonymous referees for their comments. Some of this work was carried out on the Dutch national e-infrastructure with the support of the SURF Foundation. We acknowledge Thomson-Reuters for providing us with JCR data.

**Appendix 1:** Generation of overlay maps on the basis of WoS data; for more and potentially updated information see http://www.leydesdorff.net/journals12

The procedure for generating overlay maps on the basis of journals in 2012 can be considered as an update to the one for 2011 data (Leydesdorff, Rafols, & Chen, 2013). However, clustering is now based on the algorithm provided in VOSviewer as the default. The number of journals included in 2012 is 10,936 (JCR 2012 for the Science Edition and Social Science Edition combined, October 2013). The map is based on the 10,546 journals (96.4%) in the largest component. The map can be compared with a similar map using Scopus data; see at http://www.leydesdorff.net/scopus12 and the instruction in Appendix 2 below.

*a. Generation of journal maps of science* ***without downloading*** *the sets (that is, from the retrieval at WoS):*
After entering one's search results at WoS, one can click in the left column at the bottom on "Analyze Results". In the Results Analysis that then opens as a next screen, one selects "Source Titles"; select "minimum records" (e.g., 1) and "Show" (e.g., 500). Thereafter select "All data rows"—different from the default option—and save the results to the file analyze.txt (the default). Analyze.exe (at http://www.leydesdorff.net/journals12/analyze.exe) reads this file, and write an output file (citing.txt) for VOSViewer if the file "citing.dbf" (at http://www.leydesdorff.net/journals12/citing.dbf ) is also present in the same folder.

Note that "analyze.txt" is not overwritten in case of a next run, but a new file analyze(1).txt (etc.) is generated by WoS. Rename this file before further processing. Analyze.exe overwrites files from previous runs.

*b. Using downloaded sets:*
Two programs are available online at http://www.leydesdorff.net/journals12/citing.exe and http://www.leydesdorff.net/journals12/crciting.exe, respectively, for processing a file containing downloaded data from WoS in the plain-text format (that is, with tags such as "AU" for the author field). These .exe files also require the presence of the table file at http://www.leydesdorff.net/journals12/citing.dbf in the same folder of the hard disk. (Right-click on the link for saving if necessary!) In addition to the coordinate information for the maps, the full and abbreviated titles of the journals as provided by the JCR are listed in this file. In the case of an unforeseen mismatch—for example, because of a title change—one is advised to adapt the title in the table file.

**Results**
When the programs and tables are brought into a single folder with an input file (data.txt or analyze.txt, respectively), an output file "citing.txt" is written. This file can be used as input to VOSViewer and thus be visualized as an overlay to the base map.

**Interdisciplinarity and diversity**
Both analyze.exe and citing.exe end with bringing the Rao-Stirling diversity value to the screen. This value is also written to the file "rao.txt" in the same folder. The measure provides an index between 0 and 1 of the interdisciplinarity of the set under study in terms of aggregated journal-journal citations (see also Leydesdorff *et al.*, 2013: 2578f. and in press).

**Clustering and coloring**

In the 2012 version, the results are based on the clustering algorithm of VOSviewer, and no longer on the community-finding algorithm of Blondel *et al.* (2008). However, the results of this alternate coloring scheme are available in the file citing.dbf. Using Excel, one can replace the values of the field “Cluster” with these values under the field “Blondel”. The file should be saved back thereafter as a dBase (.dbf) file with the name “citing.dbf”. (If necessary use OpenOffice or SPSS when Excel fails to do this correctly.)

**Appendix 2:** Generation of overlay maps on the basis of Scopus data; see at http://www.leydesdorff.net/scopus12

One can generate an overlay for a downloaded from Scopus as follows:

- Search in Scopus (advanced or basic); for example, using the search string 'TITLE("humanities computing") OR TITLE("computational humanities") OR TITLE("digital humanities") OR TITLE("ehumanities") OR TITLE("e-humanities")' provided 114 documents on October 8, 2013;
- Make a possible selection of records among the retrieved documents or tick "All";
- Click on "Export";
- Among the output formats, choose the "RIS format (Reference Manager, Procite, Endnote)" and "Specify fields to be exported";
- Only "Source titles" should be exported; untick all other fields;
- Click on "Export": the file "scopus.ris" can be saved; for example, for the 114 records mentioned;
- Save the file "scopus.ris" in the same folder as the routine scopus12.exe and the file with the mapping information scopus12.dbf (right-click for saving this file). One can paste different (e.g., sequential) output files of Scopus into a single file, but the routine expects a single input file with the name "scopus.ris".
- One can now run overlay.exe in that same folder; preferably from the C-prompt; (using the C-prompt, one obtains error messages);
- The file "overlay.txt" is a map file that can be opened in VOSviewer; the overlay is then visualized.
- Rao-Stirling diversity is stored in the file "rao.txt" and shown on the screen; overlay.dbf contains the information such as the number of publications in each journal;

The file scopus12.dbf contains a binary variable "status" that indicates whether the journal is included both in JCR and Scopus (status = 1) or unique to Scopus (status = 0); 10,524 journals are shared among the two databases (see Table 3 above). A complete list of journals is provided at http://www.leydesdorff.net/journals12/all_journals.xlsx .